\documentclass[conference]{IEEEtran}
\IEEEoverridecommandlockouts
\usepackage{cite}
\usepackage{amsmath,amssymb,amsfonts,bm} 
\usepackage{algorithm}
\usepackage{algorithmic}
\usepackage{graphicx}
\usepackage{textcomp}
\usepackage{xcolor}
\usepackage{subfig}
\usepackage[export]{adjustbox}
\newcommand{\herm}{\mathsf{H}}    
\newcommand{\trans}{\mathsf{T}}   
\newcommand{\Frob}{\mathsf{F}}    
\DeclareMathOperator*{\argmax}{arg\,max}
\begin{document}

\title{Antenna Coding Design for Pixel Antenna Empowered Rate-Splitting Multiple Access
\\
\thanks{This work has been supported in part by the National Nature Science Foundation of China under Grant 62571331; and in part by the Science and Technology Development Fund, Macau SAR (File/Project no. 001/2024/SKL). (\textit{Corresponding author: Yijie Mao})} 
}
\author{\IEEEauthorblockN{ Haobo Huang$^*$, Yijie Mao$^*$, Hongyu Li$^\dagger$, and Shanpu Shen$^{\ddagger}$}
\IEEEauthorblockA{$^*$School of Information Science and Technology, ShanghaiTech University, Shanghai, China \\
$^\dagger$Internet of Things Thrust, The Hong Kong University of Science and Technology (Guangzhou), Guangzhou, China \\
$^{\ddagger}$State Key Laboratory of Internet of Things for Smart City, University of Macau, Macau \\
Email: \{huanghb2025, maoyj\}@shanghaitech.edu.cn, hongyuli@hkust-gz.edu.cn, shanpushen@um.edu.mo
\vspace{0mm}
}
}

\maketitle

\begin{abstract}
This work explores the integration of pixel antennas and rate-splitting multiple access (RSMA) to enhance spectral efficiency in multi-user multiple-input single-output (MU-MISO) systems. Pixel antennas offer controllable antenna characteristics via antenna coding from the analog domain, whereas RSMA provides efficient interference management from the digital domain.
We propose a novel pixel antenna empowered RSMA transmission framework where each user employs a pixel antenna. Under imperfect channel state information at the transmitter, we formulate a joint precoding and antenna coding design problem to maximize the ergodic sum-rate. 
An alternating optimization algorithm based on the weighted minimum mean square error (WMMSE) approach and the successive exhaustive Boolean optimization (SEBO) is first developed to solve the problem. 
We then propose an efficient online antenna coder selection algorithm relying on an offline-designed codebook to reduce  computational complexity. 
Numerical results show that the proposed pixel antenna empowered RSMA significantly improves spectral efficiency compared to both RSMA with fixed antennas and space-division multiple access (SDMA) employing the same pixel antenna configuration. 
Moreover, compared to SDMA, RSMA maintains the same performance with a simpler pixel antenna configuration or a smaller codebook size.
\end{abstract}


\section{Introduction}
Recently, pixel antennas have emerged as a promising reconfigurable antenna technology to enhance wireless systems by offering additional degrees of freedom (DoF) in contrast to conventional antennas with fixed configurations (e.g., position, polarization, and radiation pattern)~\cite{pringleReconfigurableApertureAntenna2004,zhangCompactMIMOSystems2021,wongReconfigurablePixelAntennas2026}. 
The basic concept of pixel antennas is to discretize a continuous radiating surface into  small elements called pixels and enable interconnections between adjacent pixels through radio frequency (RF) switches. 
By controlling the states of RF switches, pixel antennas enable a technique named antenna coding~\cite{shenAntennaCodingEmpowered2025} to dynamically reconfigure antenna characteristics such as radiation pattern and polarization.
This helps exploit spatial multiplexing to enhance spectral efficiency for multiple-input multiple-output (MIMO)~\cite{shenAntennaCodingEmpowered2025,hanExploitingSpatialMultiplexing2026} and multi-user~\cite{liAntennaCodingDesign2025a} systems.

The system model and antenna coding technology for pixel antennas were first proposed in~\cite{shenAntennaCodingEmpowered2025}, showing that using pixel antennas can improve channel gain and capacity. 
However, this work was limited to single-user scenarios. Recent work~\cite{liAntennaCodingDesign2025a} extended the use of pixel antennas to multi-user scenarios, demonstrating that employing pixel antennas on the user side effectively improves the sum-rate performance. 
Nevertheless, this work is restricted to using space-division multiple access (SDMA) to support multiple users, and more importantly, assumes perfect channel state information at the transmitter (CSIT). 
To the best of our knowledge, the performance of pixel antennas under advanced multiple access schemes and more practical imperfect CSIT remains unexplored.
On the other hand, recent years have witnessed the development of a novel non-orthogonal multiple access scheme called rate-splitting multiple access (RSMA)~\cite{maoRateSplittingMultipleAccess2022,maoRatesplittingMultipleAccess2018,clerckxPrimerRateSplittingMultiple2023}. 
RSMA has been shown to significantly enhance spectral efficiency as well as robustness towards imperfect CSIT~\cite{joudehSumRateMaximizationLinearly2016} in multi-user systems compared to conventional multiple access techniques by enabling partial decoding of interference while treating the remaining interference as noise.

\begin{figure*}[t] 
    \centering
    \includegraphics[trim=0.6cm 0.5cm 0.8cm 0.6cm, clip, width=1\textwidth]{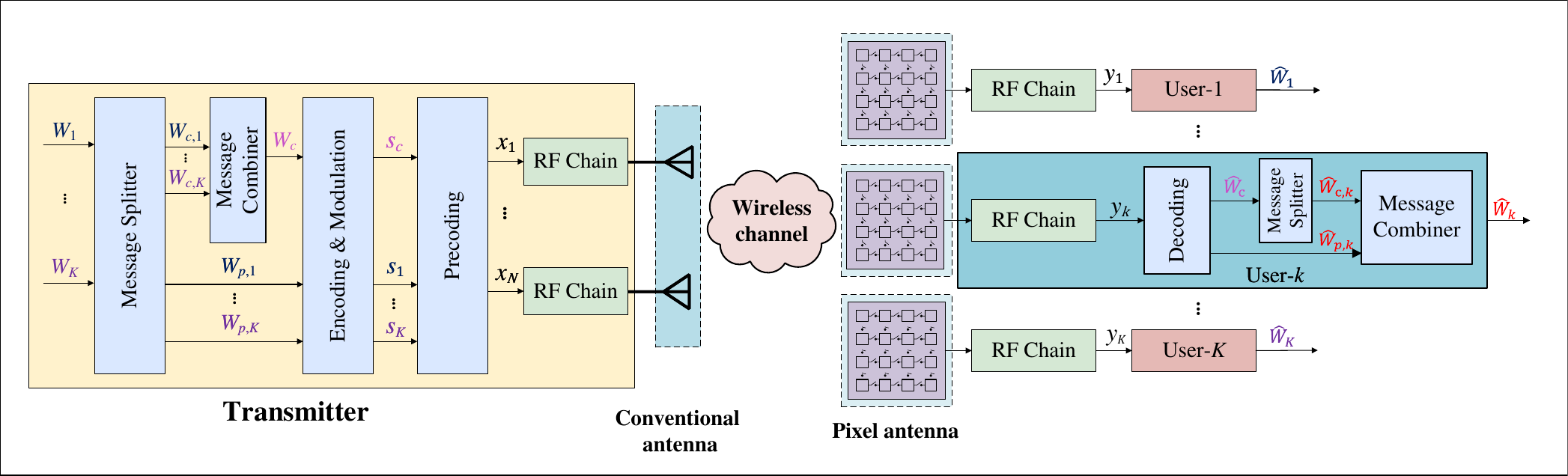} 
    \caption{The proposed pixel antenna empowered RSMA system, where the BS employs conventional antennas and each user is equipped with one pixel antenna.}
    \label{fig:1}
    \vspace{-0.35cm}
\end{figure*}

Motivated by the benefits of RSMA and the current research gap between pixel antennas and advanced multiple access schemes with imperfect CSIT, this paper explores the pixel antenna empowered RSMA transmission framework. The main contributions are summarized as follows:

\textit{First}, we propose a novel multi-user multiple-input single-output (MU-MISO) communication network, where the base station (BS) employs conventional antennas to support RSMA, while each user is equipped with one pixel antenna.
\textit{Second}, based on this model, we formulate a joint precoding and antenna coding design problem to maximize the ergodic sum-rate under imperfect CSIT. To address this problem, we develop an alternating optimization algorithm that combines the weighted minimum mean square error (WMMSE)~\cite{shiIterativelyWeightedMMSE2011} approach with successive exhaustive Boolean optimization (SEBO)~\cite{shenAntennaCodingEmpowered2025}.
\textit{Third}, to reduce computational complexity, we propose an online antenna coder selection algorithm, which relies on an offline-designed codebook. 
Numerical results demonstrate that the proposed framework significantly improves spectral efficiency compared with RSMA with conventional antennas or SDMA  with pixel antennas.
Notably, with either a low-complexity pixel antenna configuration or a smaller codebook size, the proposed framework achieves comparable or better performance than an SDMA-based system.

\section{System Model and Problem Formulation}
As illustrated in Fig.~\ref{fig:1}, we consider a downlink MU-MISO system where a BS equipped with $N$ conventional antennas communicates with $K$ users, each employing an identical pixel antenna. The user set is denoted by $\mathcal{K} = \{1, \dots, K\}$. 
Unlike conventional antennas, a pixel antenna discretizes a continuous radiation surface into pixels. Adjacent pixels are connected via RF switches, enabling a highly reconfigurable antenna that can adapt to dynamic channel conditions~\cite{shenAntennaCodingEmpowered2025}. 

\subsection{Pixel Antenna-Based Channel Model}
A pixel antenna containing $Q$ RF switches can be characterized by a $(Q+1)$-port network according to multiport network theory, with $Q$ pixel ports and one antenna port~\cite{shenAntennaCodingEmpowered2025}. The characteristic of this network is described by the impedance matrix $\mathbf{Z} = [z_{\textrm{AA}}, \mathbf{z}_{\textrm{PA}}^\trans; \mathbf{z}_{\textrm{PA}}, \mathbf{Z}_{\textrm{PP}}] \in \mathbb{C}^{(Q+1) \times (Q+1)}$, where $z_{\textrm{AA}} \in \mathbb{C}$ denotes the self-impedance of the antenna port, $\mathbf{Z}_{\textrm{PP}} \in \mathbb{C}^{Q \times Q}$ denotes the self-impedance matrix of the pixel ports, and $\mathbf{z}_{\textrm{PA}} \in \mathbb{C}^{Q \times 1}$ denotes the trans-impedance between the antenna port and the pixel ports.

The load impedance $z_{\textrm{L},q}$ of pixel port $q$ represents the state of the $q$th switch. Specifically, when the switch is on, $z_{\textrm{L},q} = 0$ corresponds to a short circuit; when the switch is off, $z_{\textrm{L},q} = \infty$ corresponds to an open circuit. 
To characterize the states of $Q$ switches, we introduce the binary vector $\mathbf{b} = [b_{1}, \dots, b_{Q}]^\trans \in \{0, 1\}^{Q\times 1}$, referred to as the antenna coder.
The load impedance matrix of the pixel ports controlled by the antenna coder $\mathbf{b}$ is $\mathbf{Z}_{\textrm{L}}(\mathbf{b}) = \textrm{diag}(z_{\textrm{L},1}, \dots, z_{\textrm{L},Q})$, where $z_{\textrm{L},q} = 0$ if $b_{q}=0$, and $z_{\textrm{L},q} = \infty$ if $b_{q}=1$.
Therefore, the currents at all ports $\mathbf{i}(\mathbf{b})\in\mathbb{C}^{(Q+1)\times 1}$ can be coded by $\mathbf{b}$ as~\cite{shenAntennaCodingEmpowered2025}
\begin{equation} \label{eq:current}
    \mathbf{i}(\mathbf{b}) = 
    \begin{bmatrix} 
    1 \\ 
    -(\mathbf{Z}_{\textrm{PP}} + \mathbf{Z}_{\textrm{L}}(\mathbf{b}))^{-1} \mathbf{z}_{\textrm{PA}} 
    \end{bmatrix} i_{\textrm{A}},
\end{equation}
where $i_{\textrm{A}}\in\mathbb{C}$ denotes the current at the antenna port. 

The pixel antenna's radiation pattern, defined as $\mathbf{e}(\mathbf{b}) \in \mathbb{C}^{2N_s \times 1}$, results from the superposition of the port radiation patterns, given by
\begin{equation}
    \mathbf{e}(\mathbf{b}) = \mathbf{E}_{\textrm{oc}}\mathbf{i}(\mathbf{b}),
\end{equation}
where $\mathbf{E}_{\textrm{oc}} = [\mathbf{e}_{\textrm{A}}^{\textrm{oc}}, \mathbf{e}_{\textrm{P},1}^{\textrm{oc}}, \dots, \mathbf{e}_{\textrm{P},Q}^{\textrm{oc}}] \in \mathbb{C}^{2N_s \times (Q+1)}$ collects the open-circuit radiation patterns of all ports, characterizing dual polarizations (azimuth and elevation) across $N_s$ spatial sampling angles.
The radiation pattern is assumed to be normalized as $\|\mathbf{e}(\mathbf{b})\|_2 = 1$.

Let $\mathbf{b}_k$ denote the antenna coder for user $k$. The corresponding port currents and normalized radiation pattern are denoted by $\mathbf{i}_k(\mathbf{b}_k)$ and $\mathbf{e}_k(\mathbf{b}_k)$, respectively.
We then obtain the beamspace representation of the channel $\mathbf{h}_k(\mathbf{b}_k)\in \mathbb{C}^{1 \times N}$ from the BS to user $k$ as~\cite{arraysignalprocessing}
\begin{equation}
    \mathbf{h}_k(\mathbf{b}_k) = \mathbf{e}_k^\trans(\mathbf{b}_k)\mathbf{H}_{\textrm{v},k}\mathbf{E}_{\textrm{T}} = \mathbf{i}_k^\trans(\mathbf{b}_k)\mathbf{E}_{\textrm{oc}}^\trans\mathbf{H}_{\textrm{v},k}\mathbf{E}_{\textrm{T}},
\end{equation}
where $\mathbf{H}_{\textrm{v},k} \in \mathbb{C}^{2N_s \times 2N_s}$ is the virtual channel matrix whose entries represent the channel gains between the angle of departure (AoD) and the angle of arrival (AoA) under dual polarizations over $N_s$ spatial angle samples, and $\mathbf{E}_{\textrm{T}} \in \mathbb{C}^{2N_s \times N}$ is the normalized radiation pattern of the transmit antennas.

We further perform singular value decomposition (SVD) for $\mathbf{E}_{\textrm{oc}}$ as $\mathbf{E}_{\textrm{oc}} = \mathbf{U}\mathbf{S}\mathbf{V}^\herm$, where $\mathbf{U}$ and $\mathbf{V}$ are semi-unitary matrices with $\mathbf{U} \in \mathbb{C}^{2N_s \times r}$ and $\mathbf{V} \in \mathbb{C}^{(Q+1) \times r}$, and $\mathbf{S} \in \mathbb{R}^{r \times r}$ collects all non-zero singular values of $\mathbf{E}_{\textrm{oc}}$. Here, $r$ is the rank of $\mathbf{E}_{\textrm{oc}}$, which is referred to as the effective aerial DoF~\cite{hanCharacteristicModeAnalysis2021}. The channel vector $\mathbf{h}_k(\mathbf{b}_k)$ is then written as
\begin{equation}
    \mathbf{h}_k(\mathbf{b}_k) = \underbrace{\mathbf{i}_k^\trans(\mathbf{b}_k) \mathbf{V}^* \mathbf{S}}_{\mathbf{w}_k^\herm(\mathbf{b}_k)} \underbrace{\mathbf{U}^\trans \mathbf{H}_{\textrm{v},k} \mathbf{E}_{\textrm{T}}}_{\mathbf{H}_{\textrm{e},k}},
    \label{eq:channel}
\end{equation}
where $\mathbf{w}_k(\mathbf{b}_k) = \mathbf{S}\mathbf{V}^\trans \mathbf{i}_k^*(\mathbf{b}_k) \in \mathbb{C}^{r \times 1}$ is named as the pattern coder for user $k$, satisfying $\|\mathbf{w}_k(\mathbf{b}_k)\|_2 = 1$, and $\mathbf{H}_{\textrm{e},k} \in \mathbb{C}^{r \times N}$ can be regarded as a channel matrix between $N$ transmit antennas and $r$ receive antennas. Hence, each pixel antenna can be equivalently viewed as a conventional array with $r$ effective antennas. 

Since the number of effective antennas $r$ is generally much smaller than spatial angle samples $N_s$, adopting  $\mathbf{H}_{\textrm{e},k}$ in \eqref{eq:channel} rather than  $\mathbf{H}_{\textrm{v},k}$ reduces both channel estimation overhead and computational complexity. Furthermore, equation~\eqref{eq:channel} reveals that the channel vector $\mathbf{h}_k(\mathbf{b}_k)$ depends on the antenna coder $\mathbf{b}_k$. Thus, the channel could be enhanced through careful antenna coding design.

In this work, we focus on a practical scenario where perfect CSIT is not available. In this case, we assume the BS has an imperfect channel estimate $\widehat{\mathbf{H}}_{\textrm{e},k}$ for user $k$ with an unknown estimation error $\widetilde{\mathbf{H}}_{\textrm{e},k}$, which follows an i.i.d. complex Gaussian distribution~\cite{joudehSumRateMaximizationLinearly2016}. Then, the real channel can be expressed as the sum of the channel estimate and its corresponding error, i.e.,
\begin{equation}
    \mathbf{H}_{\textrm{e},k} = \widehat{\mathbf{H}}_{\textrm{e},k} + \widetilde{\mathbf{H}}_{\textrm{e},k}.
\end{equation}
We assume that the average power of the estimation error scales with the transmit power $P_t$, i.e., $\mathbb{E}[\|\widetilde{\mathbf{H}}_{\textrm{e},k}\|_\Frob^2] \propto P_t^{-\alpha}$, with $\alpha \in [0,1]$ denoting the CSIT quality factor.

\subsection{RSMA Signal Model}
1-layer RSMA is employed at the BS~\cite{maoRateSplittingMultipleAccess2022}. Each user $k$'s message $W_k$ is split into two components: a common part $W_{c,k}$ and a private part $W_{p,k}$. All common parts $W_{c,1}, \dots, W_{c,K}$ are merged into a unified common message $W_c$, which is then encoded into a common data stream $s_c$ to be decoded by every user. Meanwhile, each private part $W_{p,k}$ is independently encoded into a dedicated private stream $s_k$, to be decoded solely by user $k$.

Denote the data stream and the corresponding precoding matrix as $\mathbf{s} =[s_c, s_1, \dots, s_K]^\trans$ and $\mathbf{P} = [\mathbf{p}_c, \mathbf{p}_1, \dots, \mathbf{p}_K]$, where $\mathbf{p}_c \in \mathbb{C}^{N \times 1}$ and $\mathbf{p}_k \in \mathbb{C}^{N \times 1}$ respectively denote the precoders for the common stream and user $k$'s private stream.  The transmit signal $\mathbf{x} \in \mathbb{C}^{N \times 1}$ is expressed as $\mathbf{x} = \mathbf{P}\mathbf{s}$. Assuming that $\mathbb{E}[\mathbf{s}\mathbf{s}^{\herm}] = \mathbf{I}$, the transmit power constraint $\mathbb{E}[\|\mathbf{x}\|_2^2] \le P_t$ can be equivalently expressed as $\mathrm{tr}(\mathbf{P}\mathbf{P}^{\herm}) \le P_t$.

Based on the transmit signal model and the channel modeling in~\eqref{eq:channel}, user $k$'s received signal is given by
\begin{equation}
    y_k  = \mathbf{h}_k(\mathbf{b}_k)\mathbf{x}+n_k=\mathbf{w}_k^\herm(\mathbf{b}_k) \mathbf{H}_{\textrm{e},k} \mathbf{x} + n_k,
\end{equation}
where the additive white Gaussian noise at user $k$ is denoted by $n_k$, following the distribution $\mathcal{CN}(0, \sigma^2)$. When decoding the common stream $s_c$, each user treats the interference from all private streams as noise. Assuming successful decoding of the common stream $s_c$, each user performs successive interference cancellation (SIC) to eliminate $s_c$ before decoding the corresponding private stream $s_k$.
Consequently, the signal-to-interference-plus-noise ratios (SINRs) at user $k$ for decoding $s_c$ and  $s_k$ are respectively given by
\begin{subequations}
\begin{align}
    \gamma_{c,k} &= \frac{|\mathbf{w}_k^\herm(\mathbf{b}_k) \mathbf{H}_{\textrm{e},k} \mathbf{p}_c|^2}{\sum_{j=1}^{K} |\mathbf{w}_k^\herm(\mathbf{b}_k) \mathbf{H}_{\textrm{e},k} \mathbf{p}_j|^2 + \sigma^2}, \\
    \gamma_{p,k} &= \frac{|\mathbf{w}_k^\herm(\mathbf{b}_k) \mathbf{H}_{\textrm{e},k} \mathbf{p}_k|^2}{\sum_{j \neq k} |\mathbf{w}_k^\herm(\mathbf{b}_k) \mathbf{H}_{\textrm{e},k} \mathbf{p}_j|^2 + \sigma^2}.
\end{align}
\end{subequations}
The ergodic rates of the common stream and private stream at user $k$ are defined as $\overline{R}_{c,k} = \mathbb{E}_{\mathbf{H}_{\textrm{e},k}, \widehat{\mathbf{H}}_{\textrm{e},k}} \big[ \log_2(1 + \gamma_{c,k}) \big]$ and $\overline{R}_{p,k} = \mathbb{E}_{\mathbf{H}_{\textrm{e},k}, \widehat{\mathbf{H}}_{\textrm{e},k}} \big[ \log_2(1 + \gamma_{p,k}) \big]$, respectively.

\subsection{Problem Formulation}
This work focuses on the joint optimization of the precoder matrix $\mathbf{P}$ and the antenna coder matrix $\mathbf{B} = [\mathbf{b}_1, \dots, \mathbf{b}_K]$ to maximize the ergodic sum-rate, formulated as
\begin{subequations}\label{P1}
{\fontsize{9.5pt}{11pt}\selectfont
\begin{align}
    \max_{\mathbf{P}, \mathbf{B}, \overline{R}_c, \mathbf{i}_{\textrm{A}}} \quad
    & \overline{R}_c + \sum_{k=1}^{K} \overline{R}_{p,k} \label{eq:obj} \\
    \text{s.t.} \quad
    & \overline{R}_c \le \overline{R}_{c,k},  \forall k, \label{eq:c1} \\
    & \|\mathbf{p}_c\|_2^2 + \sum_{k=1}^{K} \|\mathbf{p}_k\|_2^2 \le P_t, \label{eq:c2} \\
    & \|\mathbf{w}_k(\mathbf{b}_k)\|_2 = 1, \; \mathbf{w}_k(\mathbf{b}_k) = \mathbf{S}\mathbf{V}^\trans \mathbf{i}_k^*(\mathbf{b}_k), \forall k, \label{eq:c3} \\
    & \mathbf{i}_k(\mathbf{b}_k) = 
    \begin{bmatrix} 
    1 \\ 
    -(\mathbf{Z}_{\textrm{PP}} + \mathbf{Z}_{\textrm{L}}(\mathbf{b}_k))^{-1} \mathbf{z}_{\textrm{PA}} 
    \end{bmatrix} i_{\textrm{A},k}, \forall k, \label{eq:c4} \\
    & \mathbf{b}_k \in \{0,1\}^{Q\times 1}, \forall k, \label{eq:c5}
\end{align}
}\end{subequations}where $\mathbf{i}_{\textrm{A}} = [i_{\textrm{A},1}, \dots, i_{\textrm{A},K}]^\trans$ collects the antenna port current of each user, and $\overline{R}_c$ is an auxiliary variable introduced for the common rate. Constraint \eqref{eq:c1} guarantees the successful recovery of the common message at each user; \eqref{eq:c2} represents the transmit power limitation; \eqref{eq:c3} is the normalized pattern coder constraint for each user, which defines the pattern coder synthesizing the pixel antenna radiation pattern; \eqref{eq:c4} determines the  currents coded by the antenna coder; and  \eqref{eq:c5} restricts $\mathbf{b}_k$ to binary variables.

\section{Proposed Optimization Framework}\label{sec:3}
To solve the stochastic non-convex problem~\eqref{P1}, we first utilize the sample average approximation (SAA) approach to convert it into a deterministic problem. Then, an alternating optimization algorithm that iteratively designs the transmit precoder matrix $\mathbf{P}$ and the antenna coder matrix $\mathbf{B}$ is proposed. To alleviate the computational overhead, we further develop a codebook-based algorithm that exploits a trained codebook.

\subsection{Sample Average Approximation}
For a given channel estimate $\widehat{\mathbf{H}}_{\textrm{e},k}$, a set of $S$ i.i.d. conditional channel error samples, denoted as $\widetilde{\mathbf{H}}_{\textrm{e},k}^{(s)}$ for $s \in \mathcal{S} = \{1, \dots, S\}$, is generated. This yields the following $S$ channel realization samples as
\begin{equation}
    \mathbf{H}_{\textrm{e},k}^{(s)} = \widehat{\mathbf{H}}_{\textrm{e},k} + \widetilde{\mathbf{H}}_{\textrm{e},k}^{(s)}, \forall s \in \{1, \dots, S\}.
\end{equation}

The average common and private rates (ARs) are defined as the conditional expectations $\widehat{R}_{c,k} = \mathbb{E}_{\mathbf{H}_{\textrm{e},k} | \widehat{\mathbf{H}}_{\textrm{e},k}}[\log_2(1 + \gamma_{c,k})]$ and $\widehat{R}_{p,k} = \mathbb{E}_{\mathbf{H}_{\textrm{e},k} | \widehat{\mathbf{H}}_{\textrm{e},k}}[\log_2(1 + \gamma_{p,k})]$. With sufficient channel samples, i.e., a sufficiently large $S$, ARs can be approximated by the sample average rates (SARs) as $\widehat{R}_{c,k}^{(S)} = \frac{1}{S} \sum_{s=1}^{S} \log_2\big(1 + \gamma_{c,k}^{(s)}\big)$ and $\widehat{R}_{p,k}^{(S)} = \frac{1}{S} \sum_{s=1}^{S} \log_2\big(1 + \gamma_{p,k}^{(s)}\big)$, respectively, where $\gamma_{c,k}^{(s)}$ and $\gamma_{p,k}^{(s)}$ are the SINRs under the $s$th channel sample $\mathbf{H}_{\textrm{e},k}^{(s)}$. Thus, the original ergodic sum-rate maximization problem is transformed into the deterministic problem for each channel estimate formulated as
\begin{subequations}\label{eq:15}
\begin{align}
    \max_{\mathbf{P}, \mathbf{B}, \widehat{R}_c^{(S)}, \mathbf{i}_{\textrm{A}}} \quad& \widehat{R}_c^{(S)} + \sum_{k=1}^{K} \widehat{R}_{p,k}^{(S)} \\
    \text{s.t.} \quad 
    & \widehat{R}_c^{(S)} \le \widehat{R}_{c,k}^{(S)},  \forall k, \label{eq:15b} \\
    & \eqref{eq:c2}\text{--}\eqref{eq:c5}.
\end{align}
\end{subequations}

\subsection{Alternating Optimization Algorithm}\label{sec:3b}
We first propose an alternating optimization algorithm to solve \eqref{eq:15}, where the  precoder matrix $\mathbf{P}$ and the antenna coder matrix $\mathbf{B}$ are iteratively updated until convergence.

When $\mathbf{B}$ is fixed, the subproblem with respect to $\mathbf{P}$ is formulated as
\begin{subequations} \label{eq:subP}
\begin{align}
    \max_{\mathbf{P}, \widehat{R}_c^{(S)}} 
    \quad & \widehat{R}_c^{(S)} + \sum_{k=1}^{K} \widehat{R}_{p,k}^{(S)} \label{eq:subP_a} \\
    \text{s.t.} \quad 
    & \eqref{eq:15b}, \eqref{eq:c2}.
\end{align}
\end{subequations}

Utilizing the WMMSE approach~\cite{joudehSumRateMaximizationLinearly2016}, for each sample $s$, we introduce equalizers $g_{c,k}^{(s)}, g_{p,k}^{(s)}$ and weights $u_{c,k}^{(s)}, u_{p,k}^{(s)}$, which are used to define the augmented weighted mean square errors (MSEs) as $\xi_{c,k}^{(s)} = u_{c,k}^{(s)}\varepsilon_{c,k}^{(s)} - \log_2(u_{c,k}^{(s)})$ and $\xi_{p,k}^{(s)} = u_{p,k}^{(s)}\varepsilon_{p,k}^{(s)} - \log_2(u_{p,k}^{(s)})$. 
Here, $\varepsilon_{c,k}^{(s)} = \mathbb{E}[|g_{c,k}^{(s)} y_k^{(s)} - s_c|^2]$ and $\varepsilon_{p,k}^{(s)} = \mathbb{E}[|g_{p,k}^{(s)} y_{p,k}^{(s)} - s_k|^2]$ are the MSEs, with $y_{p,k}^{(s)}$ denoting the received signal after SIC. 
Setting $\partial \xi_{c,k}^{(s)} / \partial g_{c,k}^{(s)} = 0$ and $\partial \xi_{c,k}^{(s)} / \partial u_{c,k}^{(s)} = 0$ (and similarly for $\xi_{p,k}^{(s)}$) yields their optimal equalizers and weights. By exploiting the rate–WMMSE relationship, substituting the optimal equalizers and weights  into the augmented weighted MSE expressions and taking the sample average yields $\widehat{\xi}_{c,k}^{(S)\star} = 1 - \widehat{R}_{c,k}^{(S)}$ and $\widehat{\xi}_{p,k}^{(S)\star} = 1 - \widehat{R}_{p,k}^{(S)}$.
Thus, \eqref{eq:subP} is converted into the following convex problem:
\begin{subequations} \label{eq:subP_WMMSE}
\begin{align}
    \min_{\mathbf{P}, \widehat{\xi}_{c}^{(S)}} \quad & \widehat{\xi}_{c}^{(S)} + \sum_{k=1}^{K} \widehat{\xi}_{p,k}^{(S)} \label{eq:subP_WMMSE_a} \\
    \text{s.t.} \quad 
    & \widehat{\xi}_{c,k}^{(S)} \le \widehat{\xi}_{c}^{(S)}, \forall k, \label{eq:subP_WMMSE_b} \\
    & \eqref{eq:c2}, \label{eq:subP_WMMSE_c}
\end{align}
\end{subequations}
which can be directly solved by off-the-shelf CVX toolboxes.


When fixing $\mathbf{P}$, the problem of optimizing $\mathbf{B}$ is handled by sequentially optimizing each antenna coder $\mathbf{b}_k$. Specifically, when focusing on the design of $\mathbf{b}_k$ for user $k$, both common and private rates for all other users $j \neq k$ are constants. This allows us to design $\mathbf{b}_k$ sequentially by  writing~\eqref{eq:15b} as $\widehat{R}_c^{(S)} \le \min \big( \widehat{R}_{c,k}^{(S)}(\mathbf{b}_k), \min_{j \neq k} \widehat{R}_{c,j}^{(S)} \big)$. Consequently, the subproblem for optimizing the antenna coder $\mathbf{b}_k$ is formulated as
\begin{subequations} \label{eq:subB}
\begin{align}
    \max_{\mathbf{b}_k, i_{\textrm{A},k}} \quad & \min \Big( \widehat{R}_{c,k}^{(S)}(\mathbf{b}_k), \, \min_{j \neq k} \widehat{R}_{c,j}^{(S)} \Big) + \widehat{R}_{p,k}^{(S)}(\mathbf{b}_k)  \label{eq:subB_a} \\
    \text{s.t.} \quad 
    & \|\mathbf{w}_k(\mathbf{b}_k)\|_2 = 1, \; \mathbf{w}_k(\mathbf{b}_k) = \mathbf{S}\mathbf{V}^\trans \mathbf{i}_k^*(\mathbf{b}_k),  \label{eq:subB_b} \\
    & \mathbf{i}_k(\mathbf{b}_k) = 
    \begin{bmatrix} 
    1 \\ 
    -(\mathbf{Z}_{\textrm{PP}} + \mathbf{Z}_{\textrm{L}}(\mathbf{b}_k))^{-1} \mathbf{z}_{\textrm{PA}} \end{bmatrix} i_{\textrm{A},k},   \\
    & \mathbf{b}_k \in \{0,1\}^{Q\times 1}.
\end{align}
\end{subequations}
Note that for a given antenna coder $\mathbf{b}_k$, the optimal $i_{\textrm{A},k}$ is deterministically obtained by satisfying the normalization constraint \eqref{eq:subB_b}. Subproblem~\eqref{eq:subB} can be addressed by the SEBO algorithm~\cite{shenSuccessiveBooleanOptimization2017}. The core idea of SEBO involves two steps:

\textit{Step 1}: $\mathbf{b}_k$ is divided into $\lceil Q/J \rceil$ blocks of length $J$. By fixing other blocks, the optimal state of the current block is found by exhaustively searching over all $2^J$ combinations.
    
\textit{Step 2}: At most $J$ bits in $\mathbf{b}_k$ are randomly flipped to jump out of local optima.

Updating each antenna coder via SEBO has a search complexity of $\mathcal{O}(I 2^J)$, where $I$ and $J$ denote the number of iterations and the block size, respectively. 

By alternately updating $\mathbf{P}$ and $\mathbf{B}$, the system sum-rate is monotonically non-decreasing at each successive iteration. Since the ergodic sum-rate is strictly upper-bounded under a finite transmit power constraint, this alternating optimization algorithm is mathematically guaranteed to converge.

\begin{figure*}[t]
    \centering
    \subfloat[$5 \times 5$ pixel antennas, $M=64$.\label{fig:2a}]{
        \includegraphics[width=0.31\linewidth, valign=t]{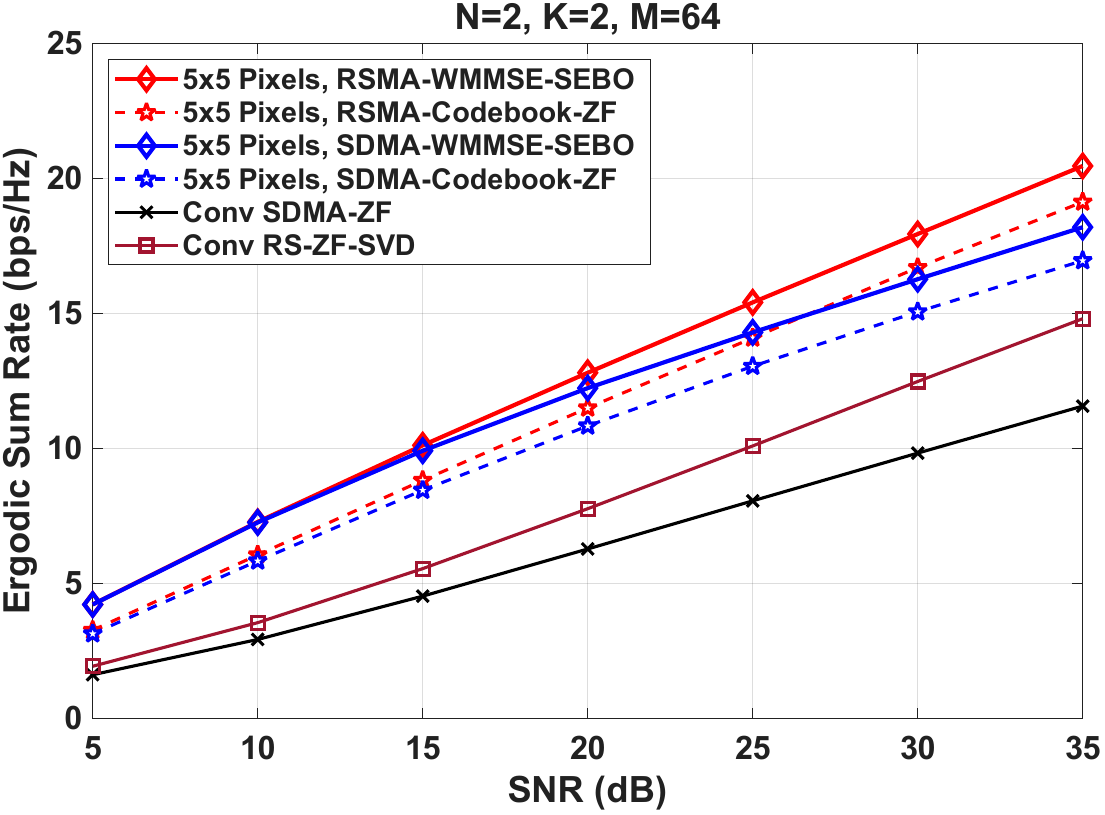}
    }
    \hfill
    \subfloat[$5 \times 5$ pixel antennas, $M=1024$.\label{fig:2b}]{
        \includegraphics[width=0.31\linewidth, valign=t]{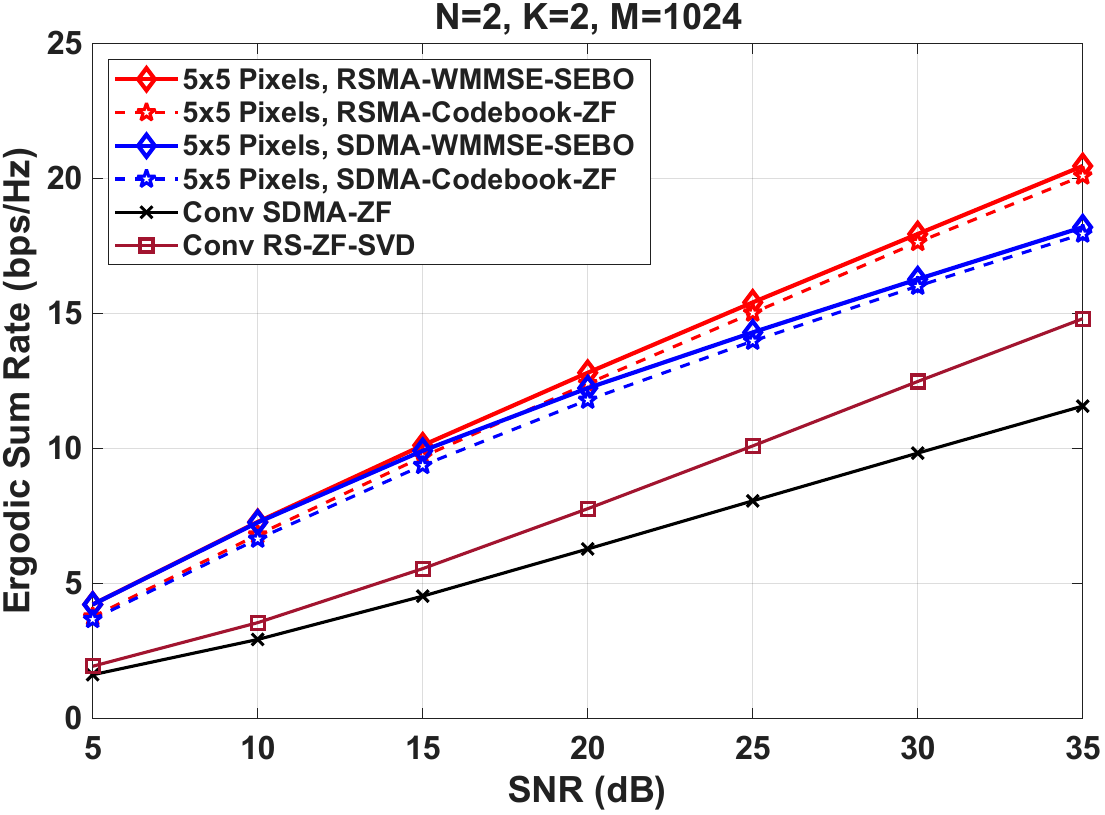}
    }
    \hfill
    \subfloat[RSMA ($3 \times 3$) vs SDMA ($5 \times 5$), $M=1024$.\label{fig:2c}]{
        \includegraphics[width=0.31\linewidth, valign=t]{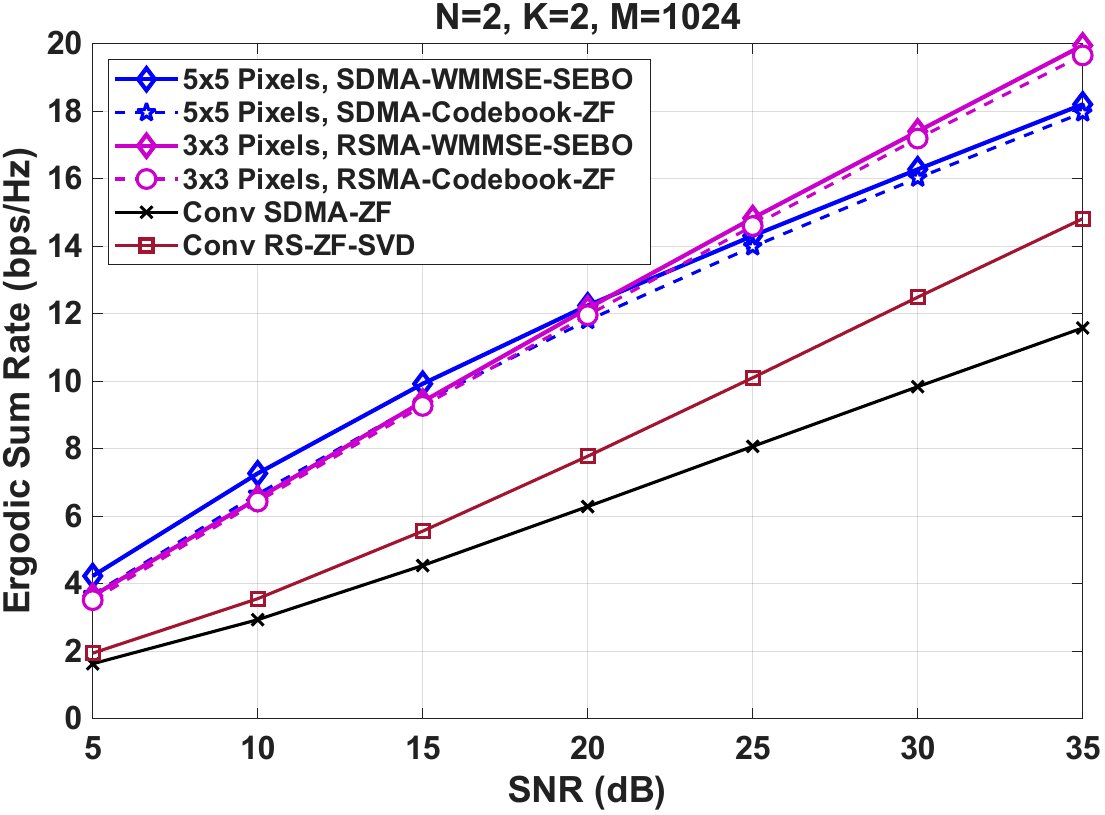}
    }
    \caption{Ergodic sum-rate comparison of different schemes under various codebook sizes and pixel antenna configurations.}
    \vspace{-0.08cm}
    \label{fig:2}
\end{figure*}

\subsection{Codebook-Based Algorithm}\label{sec:3c}
Since the alternating optimization algorithm  proposed in Section~\ref{sec:3b} entails high computational complexity by searching the entire feasible space via SEBO, we further propose a low-complexity alternative algorithm based on offline antenna codebook design and online codebook selection.

For offline codebook design, a training set $\mathcal{H} = \{\widehat{\mathbf{H}}_{\textrm{e}}^{d} \mid d = 1, \dots, D\}$ is generated under the assumption that the channel estimates of all users follow the same distribution, where $\widehat{\mathbf{H}}_{\textrm{e}}^{d} = [\widehat{\mathbf{H}}_{\textrm{e},1}^{d}, \dots, \widehat{\mathbf{H}}_{\textrm{e},K}^{d}]$ collects the channel estimates of all $K$ users for the $d$th channel realization. 
We aim to design a common codebook $\mathcal{C} = \{ \mathbf{c}_m \in \{0,1\}^{Q\times 1} \mid m = 1, \dots, M \}$ of size $M$ to maximize the global average sum-rate over the entire training set
\begin{equation} \label{eq:R_avg}
    \widehat{R}_{\textrm{avg}} = \frac{1}{|\mathcal{H}|} \sum_{m=1}^{M} \sum_{\widehat{\mathbf{H}}_{\textrm{e}}^{d} \in \mathcal{H}_m} \widehat{R}^{(S)}\big( \mathbf{c}_m \mid \widehat{\mathbf{H}}_{\textrm{e}}^{d} \big),
\end{equation}
where $\widehat{R}^{(S)}\big( \mathbf{c}_m \mid \widehat{\mathbf{H}}_{\textrm{e}}^{d} \big) = \widehat{R}_c^{(S)}(\mathbf{c}_m) + \sum_{k=1}^{K} \widehat{R}_{p,k}^{(S)}(\mathbf{c}_m)$ denotes the sample average sum-rate when all users adopt the same codeword $\mathbf{c}_m$ under $\widehat{\mathbf{H}}_{\textrm{e}}^{d}$, and $\mathcal{H}_m$ is the subset of $\mathcal{H}$ assigned to $\mathbf{c}_m$. 

Here, for each channel sample $\widehat{\mathbf{H}}_{\textrm{e}}^{d}$ and codeword $\mathbf{c}_m$, the precoder is determined by the rate-splitting zero-forcing singular value decomposition (RS-ZF-SVD) method~\cite{joudehSumRateMaximizationLinearly2016}. Specifically, for a given codeword, the common stream's precoding direction aligns with the dominant left singular vector of the concatenated coded effective channel estimate matrix, while private streams use zero-forcing (ZF) precoding. A one-dimensional linear search determines the common-private power split, with equal power allocation among private streams. In this way, we formulate the objective function in~\eqref{eq:R_avg} as a function of purely antenna coder codewords and channel estimate samples.

The offline codebook training is based on a typical Lloyd algorithm~\cite{xiaDesignAnalysisTransmitbeamforming2006}, which initializes with a randomly generated codebook $\mathcal{C}^{[0]}$. The main idea is to iteratively perform the following two steps to maximize \eqref{eq:R_avg} until convergence:

1) \textit{Training Set Partitioning}: Associated with the determined codewords $\mathcal{C}^{[l]} = \{\mathbf{c}_1^{[l]}, \dots, \mathbf{c}_M^{[l]}\}$ at the $l$th iteration, partition the training set $\mathcal{H}$ into $M$ subsets $\mathcal{H}_1^{[l]}, \dots, \mathcal{H}_M^{[l]}$, where each channel realization is assigned to the subset corresponding to the codeword that maximizes the sample average sum-rate. Assuming the codeword $\mathbf{c}_j^{[l]}$ is adopted for all users, the partition $\mathcal{H}_m^{[l]}$ is given by
{\fontsize{10pt}{12pt}\selectfont
\begin{equation} \label{eq:partition}
    \mathcal{H}_m^{[l]} = \left\{ \widehat{\mathbf{H}}_{\textrm{e}}^{d} \;\Big|\; m = \argmax_{j \in \{1,\dots,M\}} \widehat{R}^{(S)}\big( \mathbf{c}_j^{[l]} \mid \widehat{\mathbf{H}}_{\textrm{e}}^{d}\big), \; \forall d  \right\}.
\end{equation}
}
2) \textit{Updating Codebook}: For each subset $\mathcal{H}_m^{[l]}$, the centroid codeword is updated to maximize the sample average sum-rate by employing the SEBO algorithm as
{\fontsize{10pt}{12pt}\selectfont
\begin{equation} \label{eq:centroid}
    \mathbf{c}_m^{[l+1]} = \argmax_{\mathbf{c} \in \{0,1\}^{Q\times 1}} \frac{1}{|\mathcal{H}_m^{[l]}|} \sum_{\widehat{\mathbf{H}}_{\textrm{e}}^{d} \in \mathcal{H}_m^{[l]}} \widehat{R}^{(S)}\big( \mathbf{c} \mid \widehat{\mathbf{H}}_{\textrm{e}}^{d} \big).
\end{equation}}These two steps are repeated iteratively until $\widehat{R}_{\textrm{avg}}$ converges, yielding the final codebook $\mathcal{C}^\star$.

For the online antenna coder selection, the BS performs an efficient search over the offline-designed codebook $\mathcal{C}^\star$ to solve the following optimization problem:
\begin{subequations} \label{eq:online_prob}
{\fontsize{10pt}{11pt}\selectfont
\begin{align}
    \max_{\mathbf{P}, \mathbf{B}, \widehat{R}_c^{(S)}, \mathbf{i}_{\textrm{A}}} \quad & \widehat{R}_c^{(S)} + \sum_{k=1}^{K} \widehat{R}_{p,k}^{(S)} \\
    \text{s.t.} \quad 
    & \eqref{eq:15b}, \eqref{eq:c2}\text{--}\eqref{eq:c4}, \label{eq:online_prob_c1} \\
    & \mathbf{b}_k \in \mathcal{C}^\star, \forall k. \label{eq:online_prob_c2}
\end{align}
}
\end{subequations}

Initially, the antenna coders $\mathbf{B}^{[0]}$ are randomly selected from $\mathcal{C}^\star$, and the precoder $\mathbf{P}$ is determined via the RS-ZF-SVD method. Then, an iterative optimization strategy is employed, where the antenna coders for all users are updated sequentially from user $1$ to $K$ at each iteration $t$. 
When user $k$'s antenna coder is updated  by traversing the codebook, the antenna coders of other users are fixed. Specifically, the optimal coder for user $k$ at the $(t+1)$-th iteration is selected by
{\fontsize{10pt}{11pt}\selectfont
\begin{equation} \label{eq:online_update}
    \mathbf{b}_k^{[t+1]} = \argmax_{\mathbf{c}_m \in \mathcal{C}^\star} \widehat{R}^{(S)}\big( \mathbf{b}_1^{[t+1]}, \dots, \mathbf{c}_m, \dots, \mathbf{b}_K^{[t]} \mid \widehat{\mathbf{H}}_{\textrm{e}} \big),
\end{equation}
}where the antenna coder matrix for all users is given by $\mathbf{B} = [ \mathbf{b}_1^{[t+1]},\dots, \mathbf{c}_m,\dots, \mathbf{b}_K^{[t]} ]$ under the concatenated effective channel estimate matrix $\widehat{\mathbf{H}}_{\textrm{e}}$.
For each candidate codeword $\mathbf{c}_m$ tested for user $k$, $\mathbf{P}$ is instantly recalculated based on the updated coded effective channel using the RS-ZF-SVD method. 
This iterative process is repeated until the average sum-rate converges.

Compared to the $\mathcal{O}(I 2^J)$ search complexity of SEBO, the proposed online codebook search algorithm significantly reduces the antenna coder selection complexity to $\mathcal{O}(M)$.

\section{Simulation Results}
This section investigates the performance of the proposed pixel antenna empowered RSMA framework, with $N=2$ and $K=2$. The pixel antennas at both users are assumed to operate at 2.4 GHz ($\lambda = 125$ mm). Two different configurations are considered. The first configuration is a $5 \times 5$ pixel antenna whose physical aperture is $0.5\lambda \times 0.5\lambda$, comprising an antenna port and $Q=39$ pixel ports, following~\cite{shenAntennaCodingEmpowered2025}. The second configuration is a $3 \times 3$ pixel antenna whose physical aperture is $0.3\lambda \times 0.3\lambda$, comprising an antenna port and $Q=11$ pixel ports. 
The impedance parameters and open-circuit patterns are obtained via CST Studio Suite simulations. The channel estimates $\widehat{\mathbf{H}}_{\textrm{e},k}$ and channel estimation errors $\widetilde{\mathbf{H}}_{\textrm{e},k}$ are considered to follow i.i.d. complex Gaussian distributions. The number of channel samples used for SAA is $S=100$, and the ergodic sum-rate is evaluated over 1000 channel realizations. The CSIT quality scaling factor is set to $\alpha = 0.5$.

We compare the ergodic sum-rate performance of the pixel antenna empowered RSMA scheme proposed in this paper and the SDMA scheme in~\cite{liAntennaCodingDesign2025a}. Users are equipped with $5 \times 5$ pixel antennas, and the cases with codebook sizes of $M=64$ and $M=1024$ are shown in Fig.~\ref{fig:2a} and Fig.~\ref{fig:2b}, respectively. Specifically, ``WMMSE-SEBO'' refers to the alternating optimization algorithm proposed in Section~\ref{sec:3b}, while ``Codebook-ZF'' refers to the low-complexity codebook-based algorithm proposed in Section~\ref{sec:3c}.
For comparison, two baselines equipped with conventional fixed antennas at both the BS and users are considered,  denoted as ``Conv SDMA-ZF'' and ``Conv RS-ZF-SVD''. The ``Conv SDMA-ZF'' scheme employs zero-forcing precoding and equal power allocation. 
For the ``Conv RS-ZF-SVD'' scheme, the common precoder is aligned with the dominant left singular vector of the channel estimate, and ZF precoding with equal power allocation is applied to the private streams~\cite{joudehSumRateMaximizationLinearly2016}.
Simulation results demonstrate that the proposed RSMA system with user-side pixel antennas always outperforms the RSMA system with conventional antennas.
This indicates that the reconfigurability of pixel antennas provides additional DoF for the system.
Moreover, when $M=1024$, the codebook-based algorithm achieves performance close to that of the high-complexity alternating optimization algorithm, suggesting that a sufficiently large codebook allows the offline codebook-based algorithm to achieve comparable performance at a significantly reduced computational cost.
Furthermore, for both ``WMMSE-SEBO'' and ``Codebook-ZF'' algorithms, the RSMA scheme consistently achieves a better sum-rate than the SDMA scheme.
Notably, in the high SNR regime, the performance of the ``RSMA-Codebook-ZF'' scheme exceeds that of the ``SDMA-WMMSE-SEBO'' scheme, showing that RSMA can offer a better performance–complexity trade-off.
\begin{figure}[t]
    \centering
    \hspace{-0.7cm} 
    \includegraphics[width=0.7\linewidth]{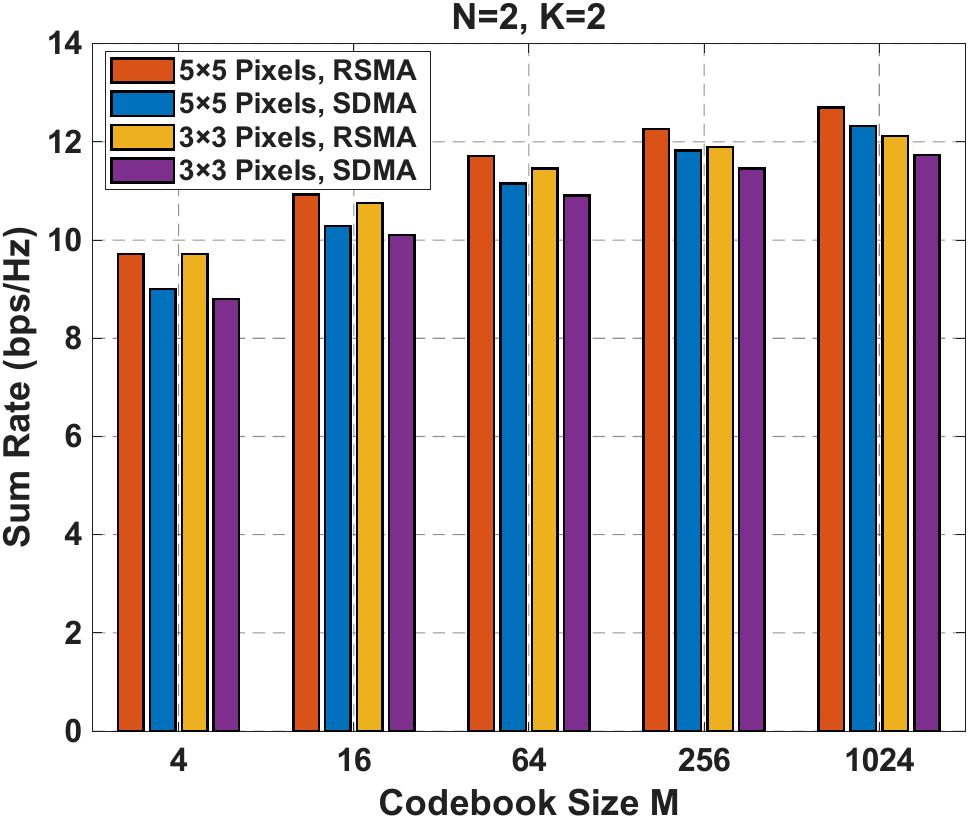}
    \caption{Ergodic sum-rate performance of different schemes with respect to the codebook size.}
    \vspace{-0.3cm}
    \label{fig:3}
\end{figure}
In Fig.~\ref{fig:2c}, we further compare the sum-rate performance between RSMA with $3 \times 3$ pixel antennas and  SDMA  with $5 \times 5$ pixel antennas at the user side under a codebook size of $M=1024$. 
The results reveal that at high SNR levels, RSMA with $3 \times 3$ pixel antennas achieves comparable or even higher performance than  SDMA  with larger $5 \times 5$ pixel antennas. 
This indicates that integrating RSMA into pixel antenna systems can maintain system performance while significantly reducing the physical aperture size and hardware complexity.

In Fig.~\ref{fig:3}, we investigate the impact of the codebook size $M$ on the sum-rate performance of the proposed codebook-based algorithm at an SNR of 20 dB, for both $5 \times 5$ and $3 \times 3$ pixel antenna configurations under RSMA and SDMA schemes. 
We observe that the ergodic sum-rate of all schemes monotonically increases with the codebook size $M$, and RSMA consistently outperforms SDMA across all codebook sizes. 
As $M$ grows, the sum-rate increment gradually decreases, reflecting a diminishing marginal gain. 
Moreover, RSMA exhibits a more evident performance advantage over SDMA as $M$ decreases.
Notably, RSMA with $3 \times 3$ pixel antennas achieves performance comparable to SDMA with $5 \times 5$ pixel antennas over the entire range of $M$. 
These results demonstrate that RSMA offers the potential to relax the hardware requirements of pixel antennas.
Furthermore, replacing SDMA with RSMA at the BS can ease the requirement of high-resolution codebooks for antenna coding at each user. 
For example, RSMA with $5 \times 5$ pixel antennas and a codebook of size $M = 64$ achieves nearly the same performance as SDMA with the same antenna configuration but $M = 256$, thereby reducing the codebook transmission overhead, storage requirements, and computational complexity.

\section{Conclusion}
This work proposes a pixel antenna empowered RSMA MU-MISO transmission framework where each user employs a pixel antenna.
Considering imperfect CSIT, we formulate a joint precoding and antenna coding design problem to maximize the ergodic sum-rate. 
We first develop an alternating optimization algorithm combining the WMMSE approach with SEBO to solve this problem. 
Additionally, we propose an online antenna coder selection algorithm based on an offline-designed codebook. 
Numerical results demonstrate that the proposed system attains significant spectral efficiency enhancement over conventional RSMA with fixed antenna configurations or SDMA with pixel antennas.
Notably, RSMA matches or even outperforms SDMA while utilizing a simpler pixel antenna configuration or a smaller codebook, thereby reducing the hardware requirements and computational complexity in pixel antenna systems.

\bibliographystyle{IEEEtran}
\bibliography{IEEEabrv, 1}

\end{document}